\documentclass[rc]{jjap3}

\title{Instability of a ferrimagnetic state 
of a frustrated $S=1/2$ Heisenberg antiferromagnet in two dimensions}
\author{Hiroki Nakano${}^{1}$
\thanks{E-mail: hnakano@sci.u-hyogo.ac.jp} 
and Toru Sakai${}^{1,2}$}
\inst{${}^{1}$Graduate School of Material Science, University of Hyogo, 
Kamigori, Hyogo 678-1297, Japan \\ 
${}^{2}$Japan Atomic Energy Agency, SPring-8, 
Sayo, Hyogo 679-5148, Japan}
\abst{To clarify the instability of the ferrimagnetism 
which is the fundamental magnetism of ferrite,  
numerical-diagonalization study is carried out for 
the two-dimensional $S=1/2$ Heisenberg antiferromagnet with frustration. 
We find that 
the ferrimagnetic ground state has the spontaneous magnetization 
in small frustration; 
due to a frustrating interaction above a specific strength, 
the spontaneous magnetization discontinuously vanishes 
so that the ferrimagnetic state appears 
only under some magnetic fields. 
We also find that, 
when the interaction is increased further, 
the ferrimagnetism disappears even under magnetic field. 
}

\begin{document}
\maketitle

%\section{Introduction}
Ferrite is a magnetic material 
that is indispensable in modern society. 
It is because this material is used 
in various industrial products including 
motors, 
generators, 
speakers, 
powder for magnetic recording, and 
magnetic heads etc. 
It is widely known that fundamental magnetism of the ferrite 
is ferrimagnetism\cite{Ichinose1,Ichinose2,Blanco,Heindl}. 
The ferrimagnetism is an important phenomenon 
that has both ferromagnetic nature
and antiferromagnetic nature at the same time.
The occurrence of ferrimagnetism 
is understood as a mathematical issue 
within the Marshall-Lieb-Mattis (MLM) theorem\cite{Marshall,Lieb_Mattis}
concerning quantum spin systems. 
A typical case showing ferrimagnetism 
is when a system includes spins of two types that 
antiferromagnetically interact between two spins of different 
types in each neighboring pair, for example, 
an ($S$, $s$)=(1, 1/2) antiferromagnetic mixed spin chain, 
in which two different spins are arranged alternately in a line 
and coupled 
by the nearest-neighbor antiferromagnetic interaction. 
The ferrimagnetic state like the above case, 
in which the spontaneous magnetization is fixed to be a
simple fraction of the saturated magnetization 
determined by the number of up spins and that of down spins 
in the state, is called the Lieb-Mattis (LM) type ferrimagnetism.  
Another example of ferrimagnetism is 
a system including single-type spins 
that are more than one in a unit cell, 
although the ferrimagnetism can appear 
even in a frustrating system including only a single spin 
within a unit cell\cite{TShimikawa_HN_s1_2,TShimikawa_HN_s1}. 

\begin{figure}
\begin{center}
\includegraphics[width=7.1cm]{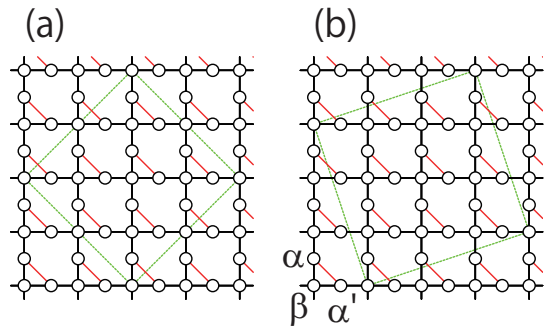}
\end{center}
\caption{(Color) Network of antiferromagnetic interactions studied in this paper. 
The black and red bonds represent $J_{1}$ and $J_{2}$ interactions. 
Green squares denote finite-size clusters of 24 and 30 sites 
in (a) and (b), respectively. 
Note that the two-dimensional network composed only of the black bonds 
is called the Lieb lattice.}
\label{fig1}
\end{figure}

The antiferromagnet on the Lieb lattice 
illustrated in Fig.~\ref{fig1} corresponds the second case, 
in which there are three spins in a unit cell. 
The MLM theorem holds 
in the Lieb-lattice antiferromagnet. 
If antiferromagnetic interactions are added to this Lieb lattice 
so that magnetic frustrations occur, 
however, the MLM theorem no longer holds. 
In this situation, the ferrimagnetic state is expected 
to become unstable. 
The problem of how the ferrimagnetism collapses owing to 
such frustrating antiferromagnetic interactions 
is an important issue 
to understand the ferrimagnetism well and 
to make ferrimagnetic materials more useful in various products. 
This problem was studied in the $S=1/2$ Heisenberg antiferromagnet 
on the spatially anisotropic kagome 
lattice\cite{2Dferri_aniso_kgm,kgm_strip}, 
where the existence of an intermediate phase 
with weak spontaneous magnetization is clarified 
between the LM type ferrimagnetic phase 
and the nonmagnetic phase 
including the isotropic kagome-lattice antiferromagnet. 
We are then faced with a question: 
is there any other different behavior of the collapse 
of the ferrimagnetism? 

Under circumstances, the purpose of this study is 
to demonstrate the existence 
of a different behavior of collapsing ferrimagnetism 
in the case of an $S=1/2$ Heisenberg antiferromagnet 
on the lattice shown in Fig.~\ref{fig1} 
to answer the above question. 
When the antiferromagnetic interactions denoted by the red bonds 
vanish, the system is unfrustrated and 
thus it certainly shows ferrimagnetism 
in the ground state. 
In this study, we examine the case 
when the red-bond interactions are switched on. 

%\section{Model and Method}
The model Hamiltonian examined in this study is given by 
${\cal H}={\cal H}_{0}+{\cal H}_{\rm Zeeman}$, where 
\begin{eqnarray}
{\cal H}_{0} &=& \sum_{i\in {\alpha},j\in {\beta}} 
J_{1} \mbox{\boldmath $S$}_{i} \cdot \mbox{\boldmath $S$}_{j} 
+
\sum_{i\in {\alpha^{\prime}},j\in {\beta}} 
J_{1} \mbox{\boldmath $S$}_{i} \cdot \mbox{\boldmath $S$}_{j} 
\nonumber \\
& & 
+
\sum_{i\in {\alpha},j\in {\alpha}^{\prime}} 
J_{2} \mbox{\boldmath $S$}_{i} \cdot \mbox{\boldmath $S$}_{j}, 
\\ 
{\cal H}_{\rm Zeeman} &=& -h \sum_{j} S_{j}^{z}. 
\label{Hamiltonian}
\end{eqnarray}
Here $\mbox{\boldmath $S$}_{i}$ denotes 
an $S=1/2$ spin operator at site $i$. 
Sublattices $\alpha$, $\alpha^{\prime}$, and $\beta$ and 
the network of antiferromagnetic interactions 
$J_{1}$ and $J_{2}$ are depicted in Fig.~\ref{fig1}. 
Here, we consider the case of isotropic interactions. 
The system size is denoted by $N_{\rm s}$. 
Energies are measured in units of $J_{1}$; 
thus, we take $J_{1}=1$ hereafter. 
We examine the properties of this model 
in the range of $J_{2}/J_{1} > 0$. 
Note that, in the case of $J_{2}=0$, 
sublattices $\alpha$ and $\alpha^{\prime}$ are combined into 
a single sublattice; the system satisfies the above conditions 
of the MLM theorem. 
Thus, ferrimagnetism of the LM type is 
exactly realized in this case. 
In the limit of $J_{2}/J_{1}\rightarrow\infty$, 
on the other hand, 
the lattice of the system is reduced to a trivial system 
composed of isolated $S=1/2$ spins and 
isolated dimers of two spins. 
Its ground state is clearly different from 
the state of the LM-type ferrimagnetism in the case of $J_{2}=0$.  
One thus finds that 
while $J_{2}$ becomes larger, 
the ground state of this system 
will change from the ferrimagnetic one in the case of $J_{2}=0$ 
to another state, which we survey here. 

\begin{figure}
\begin{center}
\includegraphics[width=7.1cm]{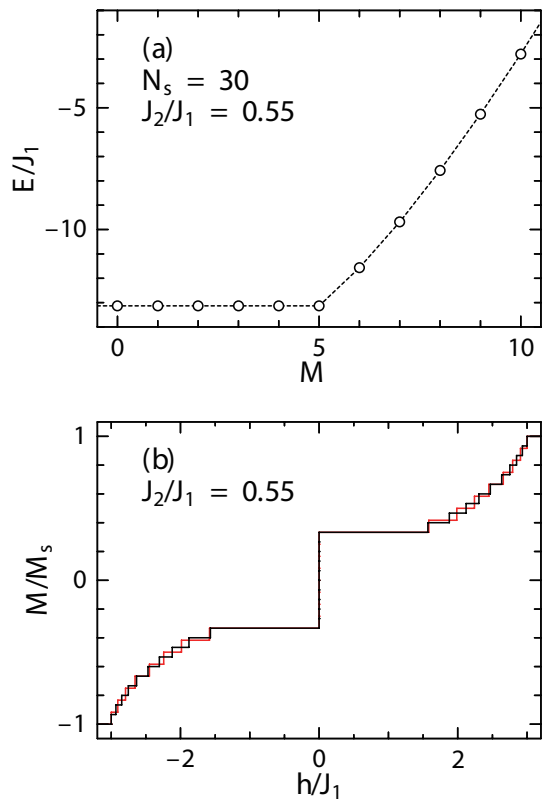}
\end{center}
\caption{(Color) Results for $J_2/J_1=0.55$. 
Lowest energy in each subspace of $M$ for the system of $N_{\rm s}=30$ 
is shown in panel (a). 
The magnetization process is depicted in panel (b); 
red and black lines represent results for $N_{\rm s}=24$ and 30, 
respectively. 
}
\label{fig2}
\end{figure}

Next, we discuss the method we use here, which is  
numerical diagonalization based on the Lanczos algorithm\cite{Lanczos}. 
It is known that this method is nonbiased 
beyond any approximations and 
reliable for many-body problems, 
which are not only localized spin systems 
such as the Heisenberg model\cite{Parkinson_Bonner,Lin_Heisenberg} 
treated in th present study 
but also strongly correlated electron systems 
including the Hubbard 
model\cite{Moreo_Hubbard,Dagotto_Hubbard,HN_2DHubbard} and 
the $t$-$J$ model\cite{Moreo_Hubbard,Ogata_tJ,
Tsunetsugu_Imada}. 
A disadvantage of this method is that the available system sizes 
are limited to being small. 
Actually, the available sizes in this method 
are much smaller than those of 
the quantum Monte Carlo simulation\cite{Miya_QMC,Todo_Kato_QMC}
and the density matrix renormalization group 
calculation\cite{White_DMRG}; however, 
it is difficult to apply both methods 
to a two-dimensional (2D) frustrated system like the present model. 
This disadvantage comes from the fact that 
the dimension of the matrix grows exponentially 
with respect to the system size. 
In this study, we treat the finite-size clusters 
depicted in Fig.~\ref{fig1} when the system sizes are 
$N_{\rm s}=24$ and 30 under the periodic boundary condition. 
Note that each of these clusters forms a regular square 
although cluster (b) is tilted from any directions 
along interaction bonds. 

We calculate the lowest energy of ${\cal H}_{0}$ 
in the subspace characterized by $\sum _j S_j^z=M$ 
by numerical diagonalizations 
based on the Lanczos algorithm and/or the Householder algorithm. 
The energy is represented by $E(N_{\rm s},M)$, 
where $M$ takes every integer up to the saturation value 
$M_{\rm s}$ ($=S N_{\rm s}$). 
We here use the normalized magnetization $m=M/M_{\rm s}$. 
Some of Lanczos diagonalizations have been carried out 
using the MPI-parallelized code, which was originally 
developed in the study of Haldane gaps\cite{HN_ATerai}. 
Note here that our program was effectively used in large-scale 
parallelized calculations\cite{kgm42_gap,HN_STodo_TSakai,kgm_1_3}. 

To obtain the magnetization process for a finite-size system, 
one finds the magnetization increase from $M$ to $M+1$ at the field 
\begin{equation}
h=E(N_{\rm s},M+1)-E(N_{\rm s},M),
\label{field_at_M}
\end{equation}
under the condition that the lowest-energy state 
with the magnetization $M$ and that with $M+1$ 
become the ground state in specific magnetic fields.  
Note here that it often happens that 
the lowest-energy state with the magnetization $M$ 
does not become the ground state in any field. 
The magnetization process in this case is determined 
around the magnetization $M$ 
by the Maxwell construction\cite{kohno_aniso2D,sakai_aniso}. 

%\section{Results and Discussions}

Now, we observe the case of $J_{2}/J_{1}=0.55$; 
results are shown in Fig. \ref{fig2}. 
Figure \ref{fig2}(a) depicts 
the lowest energy level in the subspace belonging to $M$ 
for $N_{\rm s}=30$. 
The levels for $M=0$ to $M=5$ are identical 
within the numerical accuracy. 
For $M > 5$, the energies increase with $M$. 
This behavior indicates that the spontaneous magnetization is $M=5$. 
In Fig. \ref{fig2}(b), we draw the magnetization process 
determined by eq. (\ref{field_at_M})  
in the full range from the negative to the positive saturations. 
The spontaneous magnetization $m=1/3$ appears 
and the state at $m=1/3$ shows the plateau with a large width. 
It is observed that, above $m=1/3$, the magnetization grows continuously. 
These behaviors are common 
with those of the LM ferrimagnetism 
at the unfrustrated case of $J_{\rm 2}=0$. 

\begin{figure}
\begin{center}
\includegraphics[width=7.1cm]{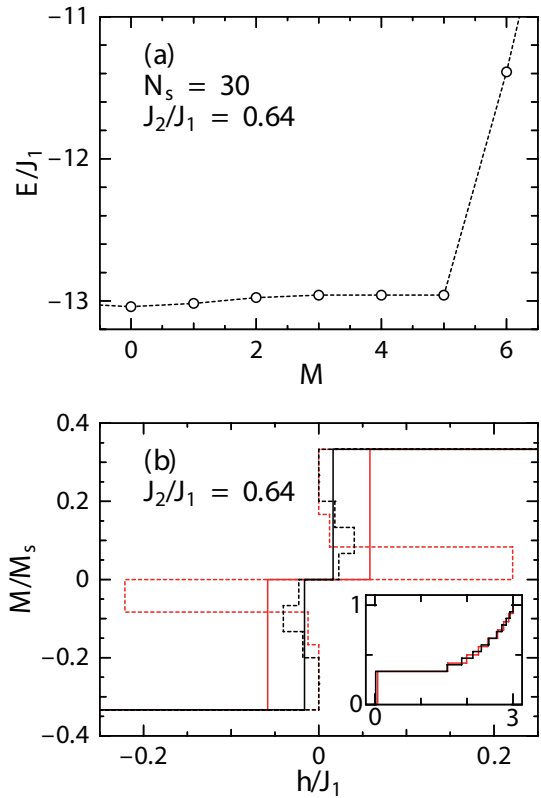}
\end{center}
\caption{(Color) Results for $J_2/J_1=0.64$. 
Lowest energy in each subspace of $M$ for the system of $N_{\rm s}=30$ 
is shown in panel (a). 
The magnetization process is depicted in panel (b); 
red and black lines represent results for $N_{\rm s}=24$ and 30, 
respectively. 
Main panel is a zoomed-in view of its inset with a wide range. 
The broken lines represent the results before the Maxwell construction 
is carried out. }
\label{fig3}
\end{figure}

Next, let us examine the case of $J_{2}/J_{1}=0.64$; 
results are shown in Fig. \ref{fig3}. 
The $M$ dependence of the lowest energy belonging to $M$ 
is different in $M < 3$ from the case of $J_{2}/J_{1}=0.55$.  
This difference affects with the disappearance 
of the spontaneous magnetization, which is shown in Fig. \ref{fig3}(b). 
This discontinuous disappearance occurs 
at $J_{2}/J_{1} \sim 0.59$ for $N_{\rm s}=24$ 
and 
at $J_{2}/J_{1} \sim 0.63$ for $N_{\rm s}=30$. 
An important point is that 
an intermediate state with smaller but nonzero spontaneous magnetizations
is absent between the $m=1/3$ state and the nonmagnetic state. 
This behavior is clearly different from the presence of such an intermediate 
state in the spatially anisotropic kagome 
lattice\cite{2Dferri_aniso_kgm,kgm_strip}. 
We speculate that this difference comes 
from the point that the competing interaction in the present model 
has a strong quantum nature localized at pairs of dimerized spins. 
The discovery of the future third case of the collapsing ferrimagnetism 
would contribute to confirm our speculation. 
Note also that 
the plateau at $m=1/3$ shows a large width. 
This suggests that the ferrimagnetic state is realized 
if external magnetic fields are added.   

\begin{figure}
\begin{center}
\includegraphics[width=7.1cm]{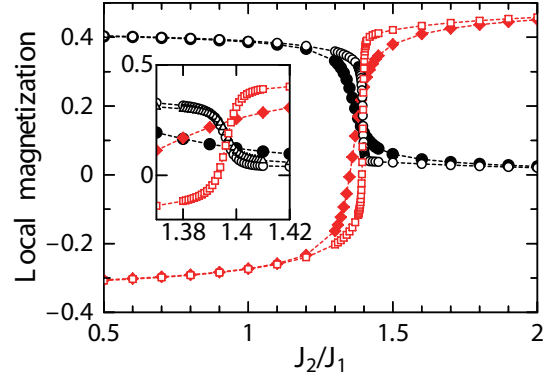}
\end{center}
\caption{(Color) Behavior of local magnetizations vs. 
the ratio of interactions $J_2/J_1$ together with 
a zoomed-in view near the transition point in inset. 
Closed circles and closed diamonds denote results for 
$\alpha$ and $\beta$ for $N_{\rm s}=24$, respectively.  
Results for $\alpha^{\prime}$ for $N_{\rm s}=24$ are 
identical those for $\alpha$ within the numerical accuracy 
because $\alpha$ and $\alpha^{\prime}$ are symmetric 
in the $N_{\rm s}=24$ cluster. 
Open circle, open triangle, and open squares represent 
results for $\alpha$, $\alpha^{\prime}$, and $\beta$ for $N_{\rm s}=30$, 
respectively. 
Due to the tilting for $N_{\rm s}=30$, $\alpha$ and $\alpha^{\prime}$ 
are not symmetric, although results of $\alpha$ and $\alpha^{\prime}$ 
for $N_{\rm s}=30$ are slightly different but very similar. 
To avoid invisibility from overlapping of symbols, 
results of $\alpha^{\prime}$ are shown only in inset. 
}
\label{fig4}
\end{figure}

\begin{figure}
\begin{center}
\includegraphics[width=7.1cm]{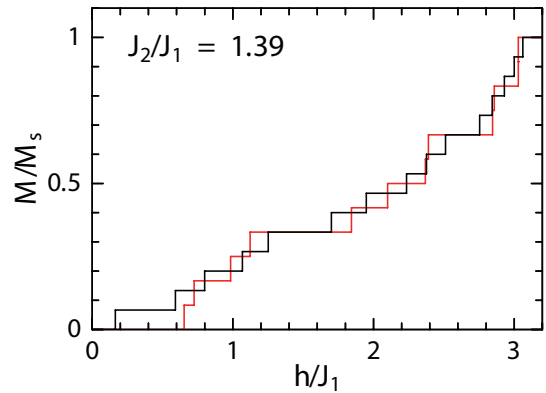}
\end{center}
\caption{(Color) Magnetization process for $J_2/J_1=1.39$. 
Red and black lines represent results for $N_{\rm s}=24$ and 30, 
respectively. }
\label{fig5}
\end{figure}

To examine 
the properties of the $m=1/3$ states in a more detailed way,  
we evaluate the local magnetization defined as 
\begin{equation}
m_{\rm LM}^{\xi} = \frac{1}{N_{\xi}} 
\sum_{j\in \xi} \langle S_j^{z} \rangle , 
\label{ave_local_mag}
\end{equation}
where $\xi$ takes $\alpha$, $\alpha^{\prime}$ and $\beta$. 
Here, the symbol $\langle {\cal O} \rangle$ 
denotes the expectation value 
of the operator ${\cal O}$ with respect 
to the lowest-energy state 
within the subspace with a fixed $M$ of interest. 
Recall here that 
the case of interest in this paper is $M=M_{\rm s}/3$. 
Here $N_{\xi}$ denotes the number of $\xi$ sites. 
Results are shown in Fig. \ref{fig4}. 
In the region of small $J_{2}/J_{1}$, 
$\alpha$ and $\alpha^{\prime}$ spins are up 
and $\beta$ spin is down, 
although each of magnetizations is slightly deviated 
from the full moment due to a quantum effect. 
This spin arrangement is a typical behavior of ferrimagnetism. 
On the other hand, 
in the region of large $J_{2}/J_{1}$, 
the magnetizations at $\alpha$ and $\alpha^{\prime}$ spins are vanishing 
and $\beta$ spin shows almost a full moment up. 
This marked change in the local magnetizations occurs 
at $J_{2}/J_{1} \sim 1.38$ for $N_{\rm s}=24$ 
and 
at $J_{2}/J_{1} \sim 1.40$ for $N_{\rm s}=30$, 
which suggests the occurrence of the phase transition 
around at $J_{2}/J_{1} \sim 1.4$. 
Therefore one finds that, 
for $J_{2}/J_{1}$ larger than this transition point, 
the ferrimagnetic state cannot be realized even under magnetic fields. 
It is unfortunately difficult to determine the transition point 
in the thermodynamic limit precisely 
only from the present two samples of small clusters. 
For the determination, 
calculations of larger clusters are required in future studies. 
Note here that 
similar observations of the local magnetizations 
were reported in Refs.~\ref{Cairo_pent_let} and \ref{Cairo_pent_full}, 
which treated the Heisenberg antiferromagnet on the Cairo-pentagon lattice 
\cite{Rousochatzakis_Cairo}, 
a 2D network obtained 
by the tiling of single-kind inequilateral pentagons.   
The same behavior of $m_{\rm LM}^{\xi}$ is also observed when 
the kagome-lattice antiferromagnet\cite{Lecheminant,Waldtmann,
KHida2001,Schulenburg,Sindzingre,kgm_ramp,kgm_ramp_PRBR} 
is distorted in the $\sqrt{3}\times\sqrt{3}$ type\cite{kgm_1_3,kgm_dist}. 
The relationship between these models should be 
examined in future studies. 
Note also that, in the present model, 
the change around the transition point seems continuous 
irrespective of whether the system size is $N_{\rm s}=24$ or 30.  
This aspect is different from the observation 
in the Cairo-pentagon-lattice 
antiferromagnet\cite{Cairo_pent_let,Cairo_pent_full}, 
where the change around the transition point seems 
continuous for $N_{\rm s}=24$ but discontinuous for $N_{\rm s}=30$. 
We speculate that whether the change is continuous or discontinuous 
in finite-size data is related to whether the number of unit cells 
in finite-size clusters is an even integer or an odd integer. 
To confirm this speculation, further investigations are required 
in future. 
It will be an unresolved question whether the transition 
is continuous or discontinuous in the thermodynamic limit. 
Figure \ref{fig5} depicts the magnetization process 
at $J_{2}/J_{1} \sim 1.39$. 
No jumps seem to appear in the process at $J_{2}/J_{1}$ corresponding 
to the transition point.  
It is unclear whether the width at $m=1/3$ survives or vanishes 
although this $m=1/3$ width at $J_{2}/J_{1} \sim 1.39$ 
is smaller than those in Figs. \ref{fig2}(b) and \ref{fig3}(b). 
Future studies would clarify 
how the magnetization process behaves 
in the vicinity of the transition point.  

%\section{Summary}
In summary, we have investigated 
how the ferrimagnetic state of the $S=1/2$ Heisenberg antiferromagnet 
on the 2D lattice collapses 
owing to magnetic frustration 
by numerical-diagonalization method. 
We capture a discontinuous vanishing 
of the spontaneous magnetization 
without intermediate phase showing spontaneous magnetizations 
that are smaller than that of the Lieb-Mattis ferrimagnetic state
when a frustrating interaction is increased.  
We also observe the disappearance 
of the ferrimagnetic state under magnetic fields 
for even larger interaction showing frustration. 
It is known that organic molecular magnets can realize 
ferrimagnetism\cite{YHosokoshi_JACS,Hosokoshi_Synth_Metals}. 
Since variety of lattice structure leading to an interaction network 
is available in such organic molecular magnets,  
the experimental confirmation might be done 
in these magnets more easily than metallic-element compounds. 
Further studies concerning instability of the ferrimagnetism would 
contribute much for our development of more stable ferrimagnetic materials. 

\acknowledgments 
This work was partly supported by JSPS KAKENHI Grant Numbers 
23340109 and 24540348. 
Nonhybrid thread-parallel calculations in numerical diagonalizations 
were based on TITPACK version 2 coded by H. Nishimori. 
Part of calculations in this study were carried out 
as an activity of a cooperative study 
in Center for Cooperative Work on Computational Science, 
University of Hyogo. 
Some of the computations were also performed using facilities 
of the Department of Simulation Science, 
National Institute for Fusion Science; 
Center for Computational Materials Science, 
Institute for Materials Research, Tohoku University; 
Supercomputer Center, Institute for Solid State Physics, 
The University of Tokyo; and Supercomputing Division, 
Information Technology Center, The University of Tokyo. 
This work was partly supported by the Strategic Programs 
for Innovative Research; the Ministry of Education, Culture, 
Sports, Science and Technology of Japan; 
and 
the Computational Materials Science Initiative, Japan. 
We also would like to express our sincere thanks 
to the staff of the Center for Computational Materials Science 
of the Institute for Materials Research, 
Tohoku University, for their continuous support 
of the SR16000 supercomputing facilities.

%\appendix
%\section{}

\end{document}